\documentclass[aps,prd,showpacs,nofootinbib,floatfix,onecolumn]{revtex4}
\usepackage{bm}
\usepackage{latexsym}
\usepackage{dcolumn}
\usepackage{amsmath,amsfonts,amssymb}
\usepackage{graphicx,epsfig}
\usepackage{color}
\usepackage{ulem}
\def\nn{\nonumber}

\def\l{\left}
\def\r{\right}
\def\DM{\mathrm{d}}
\def\lp{L_{_{\rm P}}}
\def \Rt {{}^{(3)} R}
\def \Rtt {\ {}^{(2)}R}
\newcommand{\comment}[1]{}
\reversemarginpar
\begin{document}
\title{The thermodynamic structure of Einstein tensor}

 \author{Dawood Kothawala}
 \email{ dawood.ak@gmail.com, dawood@iucaa.ernet.in}
 \affiliation{Department of Mathematics and Statistics, University of New Brunswick, Fredericton, NB, Canada E3B 5A3}

\date{\today}

\begin{abstract}
We analyze the generic structure of Einstein tensor projected onto a $2$-$D$ spacelike surface $\mathcal{S}$ defined 
by a unit timelike and spacelike vectors $\bm u$ and $\bm n$ respectively, which describe an accelerated observer (see text). Assuming that flow along $\bm u$ defines an approximate Killing vector $\bm \xi$, we then show that near the corresponding Rindler horizon, the flux $j^a=G^a_b \xi^b$ along the ingoing null geodesics $\bm k$, i.e., $\bm j \bm \cdot \bm k$,  has a natural thermodynamic interpretation. Moreover, change in cross-sectional area of the $\bm k$ congruence yields the required change in area of $\mathcal S$ under virtual displacements \textit{normal} to it. The main aim of this note is to clearly demonstrate how, and why, the content of Einstein equations under such horizon deformations, originally pointed out by Padmanabhan, is essentially different from the result of Jacobson, who employed the so called Clausius relation in an attempt to derive Einstein equations from such a Clausius relation. More specifically, 
we show how a \textit{very specific geometric term} [reminiscent of Hawking's quasi-local expression for energy of 
spheres] corresponding to change in \textit{gravitational energy} arises inevitably in the 
first law: $\DM E_{G}/ \DM \lambda \propto \int \limits^{}_{H} \DM^2 x \sqrt{\sigma} \Rtt$ (see text) -- the 
contribution of this purely geometric term would be missed in attempts to obtain area (and hence entropy) change by 
integrating the Raychaudhuri equation. 
\comment{
Moreover, we emphasize that the choice of ingoing congruence $\bm k$ is the 
correct one for discussing matter flux into a horizon; this is also in tune with the old result of T. Dray 
and G. 't Hooft [\textit{Nucl. Phys. B} \textbf{253} (1985)], from which it follows that a massless particle 
falling into a Schwarzschild black hole corresponds to a shift in the ingoing Kruskal coordinate.
}
\comment{
In fact, the well-known quasi-local nature of this energy is an 
indication that an ultra-local description of energy transfer across a horizon cross-section must be handled with 
care.
}
\end{abstract}

\maketitle
\vskip 0.5 in
\noindent
\maketitle

\section{Introduction}
The study of thermodynamic aspects of black holes over the past decades has given several insights into the nature of 
gravity as described by Einstein's General Relativity, and is expected to be a crucial link in constructing a quantum 
theory of gravity (see \cite{paddy-newinsights} for a recent review and references). In a paper more than 
a decade back \cite{jacobson-eq-of-state}, Jacobson speculated that it might be possible to 
invert the logic of the ``physical process" version of the laws of black hole mechanics, developed by Wald, and by 
applying it to local Rindler horizons, one can derive Einstein field equations from Clausius relation, 
$T \DM S = \DM E_M$, where $E_M$ is related to matter flux (and vanishes when $T_{ab}=0$). The essential new idea introduced by Jacobson was that of local Rindler horizons in a small patch 
of spacetime which can be approximated as flat once one has set the acceleration length scale appropriately. (See Appendix \ref{app:lif-conds} for an elaboration on this construction.) Einstein equations would then emerge as consistency conditions on the background. 

In a later 
paper \cite{paddy-pdv}, Padmanabhan pointed out that if one actually looks at the structure of Einstein tensor near a 
spherically symmetric horizon, it has the form $T \DM S=\DM E_G+P \DM V$, where $E_G$ is associated with horizon energy (and unlike $E_M$, $E_G \neq 0$ when $T_{ab}=0$) and $P$ with matter flux (these are defined below). In fact, the above relation has been shown to hold for a wide class of horizons, including \textit{arbitrary static horizons in Lanczos-Lovelock theory} as well. This result looks 
different from what Jacobson had started with to deduce the null-null part of Einstein equations - specifically, the energy term $E_G$ has nothing to do with $E_M$, which is more like the $P \DM V$ term but with different interpretation in terms of matter flux. So, while the Clausius relation seems to yield null-null component of Einstein equations, the Einstein tensor itself has a \textit{very different structure}. It is important to relate these results and understand where the 
difference comes from, which we intend to do in this note. 

Before proceeding, we would like to clarify an important point so as to put the analysis presented here in proper perspective. To begin with, we must mention that our main emphasis here is {\it not} to analyze pros and cons of one method over the other, but rather to clarify {\it why} they differ and to characterise the difference(s) from a physical point of view. It is indeed true that a priori there is a difference between the approaches of Jacobson and Padmanabhan; while Jacobson's analysis concerns deriving Einstein equations from Clausius relation, Padmanabhan's result demonstrates that Einstein equations on horizon is same as the first law of thermodynamics. However, once the physical content of Einstein equations has been claimed to be equivalent to a particular thermodynamic relation, one would have expected a mapping between the two results, unless there are subtle differences at a fundamental level. Indeed, the $T \DM S$ term in the thermodynamic relation is fairly unambiguous, so that the remaining terms in the equations must correspond in some manner. If they do not, then it implies that there is a difference at a conceptual level, which is what we shall show in this note. We shall show that \textit{the difference arises in the 
particular manner in which matter fluxes across the horizon are treated.} Specifically, Padmanabhan's result arises
due to deformations of the future horizon \textit{normal} to itself, generated by ingoing null geodesic 
congruences, and this yields the force term $P \DM V$ in the final result. We discuss in some detail the resulting 
difference in physical interpretations. Furthermore, we show that the additional term $\DM E_G$ is essentially the 
change in quasi-local energy associated with the horizon $2$-surface, and is related to horizon topology; more 
precisely, we show that $\DM E_G/ \DM \lambda \propto \int \DM^2 x \ \sqrt{\sigma} \Rtt$. To summarize, we 
shall do the following in this note:

\begin{enumerate}

\item Clarify the role of horizon deformations to be considered in a Rindler patch when matter crosses the future Rindler horizon of the observer.

\item Clarify the differences between the ``heat flux" term of Jacobson, and the ``$P \DM V$" term of Padmanabhan, 
and highlight the physical implications.

\item Indicate clearly that the change in area of a horizon cross-section is determined by the expansion (and not its first derivative) of the ingoing null congruence normalised to have unit Killing energy. \comment{Using the outgoing congruence, \textit{normalised in this same way}, also gives the same result, although we do not use them since these have divergent components in the inertial frame (see below).}

\item Give an explicit expression for the expansion $\theta$ for the congruence mentioned in the previous point, in terms of combination of curvature tensor components (see Eq.~(\ref{eq:riemm-area-change}) below), and compare with the corresponding combination occuring in the Raychaudhuri equation. In particular, {\it the area change involves not just the Ricci tensor, but also the Riemann tensor} -- a point which is of relevance in the context of deriving field equations from thermodynamics.

\item Exhibit explicitly the ``thermodynamic" structure of Einstein tensor and show that there is a term corresponding 
to quasi-local energy of the horizon, which must be separately accounted for when considering energy flow across the horizon.

\item Show that, when using the Raychaudhuri equation with our prescribed null congruence,  the $O(\lambda)$ term does not vindicate or necessitate setting the expansion to zero.


\end{enumerate}

We shall address all the above points in the following sections. To avoid distraction from the main points, we have relegated most of the mathematical details to appendices. Before proceeding, let us also clarify the restrictions on the local frame of the accelerated observer that we shall impose. The most important restriction is that of staticity; that is, we shall assume that, in the local coordinates near the observer worldline, one can define an approximate timelike Killing vector field. Consequently, the near horizon geometry is assumed to be static. For static spacetimes, we shall use, for the near horizon metric, the form: $\DM s^2 = -N^2 \DM t^2 + \DM z^2 + \sigma_{AB} \DM y^A \DM y^B$, with the Taylor expansions for 
$N$ and $\sigma_{AB}$ derived by Visser et al \cite{visser}. As our discussion will make clear, the above form of 
metric is just a good, convenient parametrization -- the final results are of course stated in a manifestly tensorial 
form. The only crucial input is staticity, which requires a satisfactory notion of a timelike Killing vector which is 
hypersurface orthogonal, and a spacelike surface whose unit normal points in the direction of acceleration. 
\comment{Also, for greater clarity and notational convenience, we use boldface subscripts for contraction on 
corresponding vectors; for e.g., $R_{\bm \xi \bm k} \equiv R_{ab} \xi^{a} k^{b} $.}

\section{The null basis near a horizon}
Let us concentrate on the future horizon $\mathcal{H}$ of the right Rindler wedge, which is generated by outgoing 
null rays. The most natural transverse null vector for $\mathcal{H}$ is therefore defined by affinely parametrized \cite{conf-proc}
ingoing null geodesics, 
$\bm k$, and can be chosen to be: $\bm k = N^{-1} (\bm u - \bm n)$. Here, $N=\sqrt{-\xi^2}, \bm u = \bm \xi/N$, and 
existence of a local timelike Killing field $\bm \xi$ (which generates local Lorentz boosts) is assumed. Also, 
$\bm n$ is the unit normal in the direction of acceleration of $\bm u$. The existence of $\bm \xi$ is also assumed in the work of Jacobson \cite{jacobson-eq-of-state}, and without this no further progress can be made.


The choice of normalization is such that $\bm k \cdot \bm \xi = -1$, implying that $\bm k$ has unit Killing energy. 
It must be also noted that the corresponding outgoing null rays are given by $\bm l = N^{-1} (\bm u + \bm n)$; we note 
that, $N^2 \bm l \rightarrow \bm \xi$ on $\mathcal{H}$. More precisely, $N^2 \bm l$ become tangent to horizon 
generators [the vector $\bm l$ itself does not, since $\bm l \cdot \bm \xi = -1$ by construction]. The standard 
Rindler transformations in the local inertial fram (LIF) has an additional parameter $\kappa$ which characterizes 
orbits of Lorentz boosts and generates constant acceleration trajectories. In inertial coordinates $(T,X,Y,Z)$, we 
have $\bm k=\kappa^{-1} (X+T)^{-1} \l( \bm \partial_T - \bm \partial_X \r) \rightarrow (2 \kappa X)^{-1} \l( \bm \partial_T - \bm \partial_X \r)$ on $\mathcal{H}$, i.e., $T=X$. In fact, we could as well have used $\bm l$ below for 
discussion without any change in the final result, but this would be a weird thing to do since these geodesics behave 
badly near the future horizon. Specifically, $\bm l=\kappa^{-1} (X-T)^{-1} \l( \bm \partial_T + \bm \partial_X \r)$, 
so the components in the locally inertial coordinates blow up at $X=T$.
\comment{
\footnote{
Actually, the vector $\bm k$ we are dealing with is the so called RIGGING vector field, which defines a natural 
projector onto the horizon. [See also the \textit{Living Reviews} article by Ashtekar et. al. on isolated/dynamical 
horizons.] 
}
}
%
%

We shall now demonstrate that the expansion of $\bm k$ (or $\bm l$) governs the changes in 
cross sectional area of the horizon. To exhibit the result for both $\bm k$ and $\bm l$ simultaneously, we 
write $\bm k_{\epsilon} =  N^{-1} \l( \bm u - \epsilon \bm n \r)$, where $\epsilon=+1$ corresponds to $\bm k$ and $\epsilon=-1$ 
to $\bm l$. First, note that, in terms of the covariant derivative ${}^{(3)}D$ compatible with $t=$constant 
hypersurface, we have
\begin{eqnarray}
{}^{(3)}D \cdot \bm n &=& \l( g^{ab} + u^a u^b \r) \nabla_a n_b
\nn \\
&=& \nabla \cdot \bm n - \bm n \cdot \bm a
\end{eqnarray}
Now evaluate
\begin{eqnarray}
\nabla \cdot \bm k_{\epsilon} &=& -\epsilon N^{-1} \nabla \cdot \bm n + N \bm k_{\epsilon} \cdot \nabla N^{-1}
\nn \\
&=& -\epsilon \frac{1}{N} \; {}^{(3)}D \cdot \bm n - \epsilon \frac{\bm a \cdot \bm n}{N} - 
\frac{\bm k_{\epsilon} \cdot \nabla N}{N}
\nn \\
&=& - \epsilon \frac{1}{N} \; {}^{(3)}D \cdot \bm n
\end{eqnarray}
where we have noted that $\bm a=\nabla N/N$, so that the last two terms in second equality cancel. Now, since the 
$t=$constant metric is $\DM z^2 + \sigma_{AB} \DM y^A \DM y^B$, the $\bm n={\bm \partial}_z$ congruence is an affinely parametrized 
geodesic congruence, and therefore we can use the standard interpretation of ${}^{(3)}D \cdot \bm n$ in terms of 
fractional rate of change in ``volume" of $z=$constant surfaces, which corresponds to the $2$-$D$ manifold described 
by the metric $\sigma_{AB}$. Hence, we finally get
\begin{eqnarray}
\nabla \cdot \bm k_{\epsilon} = - \epsilon \frac{1}{N} \partial_z \ln \sqrt{\sigma} 
= - \epsilon \frac{\DM ~}{\DM \lambda} \ln \sqrt{\sigma}
\label{eq:NEW-exp-area-change}
\end{eqnarray}
which is the desired result. (We have used $N \DM z = \DM \lambda$ in arriving at the second equality, see Appendix 
\ref{app:wald} for details.) This straightforward evaluation should leave no doubt as to how the change in 
cross sectional area of the horizon is actually described by considering ingoing (or outgoing) null geodesic 
congruences, constructed in the manner we have described. In fact, the choice of ingoing congruence $\bm k$ is 
also strengthened by some old results due to T. Dray and G. 't Hooft \cite{dray-thooft}, which clearly shows that a 
massless particle falling into a Schwarzschild black hole corresponds to a shift in the \textit{ingoing} Kruskal 
coordinate, the shift being proportional to the particle energy.
\section{The thermodynamic structure of Einstein tensor}
We shall now analyze the near-horizon form of Einstein tensor, and reveal how its thermodynamic structure emerges. 
Before proceeding, however, we wish to clarify an important point concerning the variations we shall be considering. We shall base our discussion on the ingoing null geodesics $\bm k$ of the previous section, satisfying 
$\bm k \cdot \bm \xi = -1$. As should be evident from comments in the previous section, \textit{the entire analysis 
can be 
repeated in a straightforward manner using outgoing null geodesics $\bm l$ satisfying $\bm l \cdot \bm \xi =-1$}; the 
only difference is the change in sign in $\bm n$ at various intermediate steps, while 
\textit{the final result remains unchanged}. The reason for using ingoing null geodesics $\bm k$, as mentioned above, 
is that these have components which are well behaved at the future horizon in the locally inertial coordinates; the 
only crucial thing is the normalization based on unit Killing energy.

We begin with the following (exact) identity [see Appendix \ref{app:gauss-codazzi-rel} for a proof]:
\begin{eqnarray}
G_{ab} g^{\perp ab} = 2 \l( R_{ab}\xi^ak^b\comment{R_{\bm \xi \bm k}} - R_{abcd}u^an^bu^cn^d\comment{R_{\bm u \bm n \bm u \bm n}} \r) &-& \Rtt
\nn \\
&-& N^2 R_{ab}k^ak^b\comment{R_{\bm k \bm k}}  + \Pi[K,k]
\label{eq:eins-struc}
\end{eqnarray}
where $g^{\perp}_{ab}=-u_a u_b + n_a n_b$ is the metric on the surface orthogonal to the horizon, 
and $\Pi[K,k]=f(k) -  f(K) - \phi(K)$, with $f(K)=K^2 - K_{\mu \nu}^2$ (similarly for $f(k)$), and 
$\phi(K)=n^{\mu} n^{\rho} \l( K^{\nu}_{\mu} K_{\nu \rho} - K K_{\mu \rho} \r)$. Here, $K_{\mu \nu}$ and $k_{AB}$ are extrinsic curvatures of level surfaces of $\bm u$ embedded in $4$-$D$ spacetime, and of $\bm n$ embedded in the resultant 
$3$-$D$ space, respectively. Note that the above expression is true for an arbitrary spacetime without any geometric 
constraints imposed so far. 
\footnote{In particular, for a flat $3$-$D$ space in a flat $4$-$D$ spacetime, one obtains $\Rtt = f(k)$, which is 
essentially the content of Gauss's \textit{Theorema Egregium}.}

We shall now impose the condition of staticity, that is, we shall require that the near horizon geometry, to a sufficient approximation, has a local timelike Killing vector field. In that case, we can show that (see Appendix \ref{app:riemm-area-change}), on the horizon $z \rightarrow 0$:
\begin{eqnarray}
R_{ab}\xi^ak^b - R_{abcd}u^an^bu^cn^d\comment{R_{\bm \xi \bm k} - R_{\bm u \bm n \bm u \bm n}} = \kappa \frac{\DM ~}{\DM \lambda} \ln \sqrt{\sigma}
\label{eq:riemm-area-change}
\end{eqnarray}
The above equation gives the derivative of area (rather than it's second derivative) in terms of curvature components, and deserves several comments, which we list below: 
\begin{itemize}
\item It clearly shows that the change in cross sectional area (obtained by integrating $\sqrt{\sigma}$ over 
transverse coordinates) of the $\bm k$ (or the $\bm l$) congruence (normalized so as to have unit Killing energy), on 
a cross section of $\mathcal{H}$, depends on a very different combination of {\it Riemann tensor} components than 
the one occurring in the Raychaudhuri equation [which only involves Ricci tensor, $R_{ab}k^ak^b\comment{R_{\bm k \bm k}}$]. 

\item Raychaudhuri equation gives \textit{second} derivative of area and our analysis above shows 
that ``integrating" it naively to obtain the first derivative will, in general, be tricky. Indeed, the null-null component does not 
appear in the above equation at all! In section \ref{sec:raych-jacobson}, we shall present an analysis a la Jacobson 
using Raychaudhuri equation, which should clarify further what is going on here. 

(This and the previous comment are important particularly when we consider Jacobson's argument and compare it with our 
result, see section \ref{sec:raych-jacobson}.)

\item The appearance of $R_{abcd}u^an^bu^cn^d\comment{R_{\bm u \bm n \bm u \bm n}}$ also must be highlighted; one could 
have simply ignored this term by demanding it to be small, and calling this demand a further restriction on the 
definition of a local Rindler horizon. This, however, would be adhoc, since for Schwarzschild horizon, it 
involves $\partial_r^2 (1-2M/r)$. Indeed, as is evident from above, there is actually no need to throw away this term, 
since it occurs in just the right combination in Einstein tensor so as to give the change in area correctly. 

\item Even if we did throw away the $R_{abcd}u^an^bu^cn^d\comment{R_{\bm u \bm n \bm u \bm n}}$ term,  we are left with $R_{ab}\xi^ak^b\comment{R_{\bm \xi \bm k}}$ which has nothing to do with the null-null component of Ricci [recall 
that $\bm k \cdot \bm \xi = -1$].

\end{itemize}

Proceeding to the main analyis, note that if $R_{ab}k^ak^b\comment{R_{\bm k \bm k}}$ [and hence 
$G_{ab}k^ak^b\comment{G_{\bm k \bm k}}$] is finite on the horizon, then the corresponding term on RHS of 
Eq.~(\ref{eq:eins-struc}) is $O(z^2)$. Also, $\Pi[K,k]$ is ignorable because it is $O(z^2)$. This comes about as follows: $K_{\mu \nu}$ is zero due to staticity. On the other hand, $k_{AB} \propto \partial_z \sigma_{AB}$ is $O(z)$ since, from the Taylor expansion of area, $\sigma_{AB}=$ ($z$-${\rm independent~part}$) $+O(z^2)$ (see Ref. \cite{visser}). Since $\Pi[K,k]$ is quadratic in $k_{AB}$, it is $O(z^2)$. So we finally 
obtain:

\begin{eqnarray}
P \sqrt \sigma = \frac{\kappa}{2 \pi} \frac{\DM ~}{\DM \lambda} \l( \frac{1}{4} \sqrt{\sigma} \r) - 
\frac{1}{16 \pi} \Rtt \sqrt{\sigma}
\label{eq:eq6}
\end{eqnarray}
where we have defined $P = (1/2) T_{ab} g^{\perp ab}$. The differential version of the above equation (multiplying it by $\DM \lambda$)
yields Padmanabhan's result: 
\begin{eqnarray}
P \DM V = T \DM S - \DM E_G
\end{eqnarray}

Having established the above relation, we can ask how general it is. It might seem that the result is very specific to Einstein gravity since in arriving at it, we used Eq.~(\ref{eq:riemm-area-change}) for change of area, and in Einstein gravity horizon entropy is proportional to area. We could therefore relate entropy change to area change and derive the result. However, when one goes beyond Einstein theory, entropy is no longer proportional to area but is instead given by Wald entropy. It is therefore quite a non-trivial fact that exactly the same result can be proved for a much larger class of lagrangians -- the so called Lanczos-Lovelock (LL) lagrangians -- for which the horizon entropy is given by a non-trivial function of area. Once again, we find that the near-horizon structure of field equations for LL actions can be cast in the form: 
\begin{eqnarray}
P \DM V = T \DM S_{LL} - \DM E_{(G) LL}
\end{eqnarray}
and the resulting expressions for $S_{LL}$ and $E_{(G) LL}$ turn out to be \cite{thermod-static}:
\begin{eqnarray}
S_{\mathrm{LL}} \propto \int \sqrt{\sigma} L_{m-1}^{(D-2)}
\\
(\DM E_{G}/\DM \lambda)_{\mathrm{LL}} \propto \int \sqrt{\sigma} L_{m}^{(D-2)}
\end{eqnarray}
We see that $S$ is precisely the Wald entropy, whereas $E_G$ gives the correct expression for quasilocal energy when applied to known black hole solutions. \footnote{A general definition for quasilocal energy, such as Hawking's definition for Einstein theory, is not available for the LL actions; in fact, ours can be taken as a natural generalization of Hawking's quasilocal energy for LL actions.}

Before turning to Raychaudhuri equation, let us make another relevant comment: It is easy to see, from our 
definitions, that $G_{ab} g^{\perp ab} = -2 G_{ab}\xi^ak^b\comment{G_{\bm k \bm \xi}} + 
N^2 G_{ab}k^ak^b\comment{G_{\bm k \bm k}}$. Therefore, 
provided $G_{ab}k^ak^b\comment{G_{\bm k \bm k}}$ is finite on the horizon, one obtains, in the limit $N \rightarrow 0$: 
$G_{ab}\xi^ak^b\comment{G_{\bm k \bm \xi}} \rightarrow - (1/2) G_{ab} g^{\perp ab}$ (which, incidentally, is the so 
called \textit{work function}, $W$, defined by Hayward in the context of \textit{spherically symmetric}, dynamical 
horizons \cite{hayward}; note that we have \textit{not} assumed spherical symmetry to obtain Eq.~(\ref{eq:eq6})). On 
the horizon, we therefore have natural interpretation for this term as the force acting on the horizon in the direction 
defined by $\bm k$. Let us 
also mention its form for an ideal fluid, described by $T_{ab}=\rho_0 v_av_b + p_0(g_{ab}+v_av_b)$, where $v^a$ is the 
fluid $4$-velocity, and we assume for simplicity that it lies only in the $\bm u$--$\bm n$ plain. Then, a trivial 
calculation shows that $T_{ab}u^au^b=\gamma_{rel}^2 (\rho_0+p_0v_{rel}^2)$ and 
$T_{ab}n^an^b=\gamma_{rel}^2 (p_0+\rho_0v_{rel}^2)$, where $\gamma_{rel}=-\bm u \cdot \bm v$. We then immediately 
obtain $P = (1/2) T_{ab} g^{\perp ab}=(p_0-\rho_0)/2$.

It is also instructive to compare this analysis with the one given by Jacobson, in which case the most natural 
starting point would be the Raychaudhuri equation. We do this in section \ref{sec:raych-jacobson}. We shall show 
that, for our $\bm k$ (or $\bm l$) congruence, the starting assumption of equating $T \DM S$ with matter flux gives, 
at $O(\lambda^0)$, a relation which is inconsistent with the algebraic identity obtained in this section. However, if 
one makes further approximations and ignore certain terms, then we do recover the null-null part of Einstein equations at $O(\lambda)$, although in a manner completely different from Jacobson's, since our analysis is not based on the null generators. Most importantly, we do not require the vanishing of expansion of the null congruence at all. Before proceeding, we must emphasize that, in the next section, we shall be trying to follow Jacobson's reasoning \textit{in our setup}; the final results and implications must, of course, be interpreted keeping this in mind. Needless to say, our main emphasis is towards trying to understand why there are differences between the work and energy terms in the two approaches; the answer, as we hope this note would make evident, lies in different ways of treating fluxes across the horizon.
\vspace{0.2in}
\section{Analysis based on Raychaudhuri equation} \label{sec:raych-jacobson}

In this section, we turn to Raychaudhuri equation, in an attempt to understand better the difference between 
above result and 
Jacobson's derivation of the null-null component of the field equations. To do so, we repeat Jacobson's analysis using 
the $\bm k$ congruence; this should indicate where the difference lies. Once again, it is worth emphasizing that we 
will obtain the same results upon using the outgoing $\bm l$ congruence of unit Killing energy. As we have shown above, 
the Einstein tensor on the whole has a much richer structure due to the presence of the $\Rtt$ term, which we would 
want to explore further. Unfortunately, Raychaudhuri equation, as we will see, has nothing much to say about this term, 
but our analysis will {shed some light on the role of certain assumptions in Jacobson's derivation, and also the differences between the work term as well as horizon energy.}

Start with the equation defining variation of area in terms of expansion $\theta$ of a congruence of ingoing 
null geodesics w.r.t. the affine parameter $\lambda$ along $\bm k$ (see Appendix \ref{app:wald} for more 
details). Assuming that entropy is proportional to area, this gives:
\begin{eqnarray}
T_H \DM S = \alpha^{-1} \int \theta \; \DM \Sigma \; \DM \lambda
\end{eqnarray}
where $\alpha = (8 \pi c \lp^2/ \hbar) / \kappa$, and the integration is over the null 3-surface generated by the cross-section of a bundle of ingoing null geodesics $\bm k$ across an affine distance $\lambda$. The horizon is at $\lambda=0$. Now expand $\theta$
\begin{eqnarray}
\theta(\lambda) = \theta(0) + \dot \theta (0) \lambda + \frac{1}{2} \ddot \theta (0) \lambda^2 + O(\lambda^3)
\end{eqnarray}
in obvious notation. Now we can use Raychaudhri equation to substitute for the first derivative 
of $\theta$ {\it evaluated at $\lambda=0$}. That is,
\begin{eqnarray}
\dot \theta (0) = - \frac{1}{2} \theta^2(0) - \l[ R_{ab}k^ak^b\comment{R_{\bm k \bm k}} \r]_{\lambda=0}
\end{eqnarray}
where \comment{$R_{\bm k \bm k}=R_{ab} k^a k^b$ and }we have ignored shear and rotation for the time being (which is 
also an assumption in Jacobson's work). 

Now consider the heat flux through $\DM \Sigma \; \DM \lambda$:
\begin{eqnarray}
\DM Q = \int T_{ab} \xi^a k^b \; \DM \Sigma \; \DM \lambda \comment{= \int T_{\bm \xi \bm k} \; \DM \Sigma \; 
\DM \lambda}  
\end{eqnarray}
for which a similar expansion gives:
\begin{eqnarray}
T_{ab}\xi^ak^b\comment{T_{\bm \xi \bm k}} = \l[ T_{ab}\xi^ak^b\comment{T_{\bm \xi \bm k}} \r]_{\lambda=0} + 
\lambda \; \l[ \frac{\DM}{\DM \lambda} T_{ab}\xi^ak^b\comment{T_{\bm \xi \bm k}} \r]_{\lambda=0} + O(\lambda^2)
\end{eqnarray}
Following Jacobson, we now impose the Clausius relation, $T \DM S = \DM Q$ and equate equal powers of 
$\lambda$ on both sides. That is,
\begin{eqnarray}
\alpha^{-1} \int \l[ \theta(0) + \dot \theta (0) \lambda + \frac{1}{2} \ddot \theta (0) \lambda^2 + O(\lambda^3) \r]
 \; \DM \Sigma \; \DM \lambda = \int \l[  \l[ T_{ab}\xi^ak^b\comment{T_{\bm \xi \bm k}} \r]_{\lambda=0} + 
\lambda \; \l[ \frac{\DM}{\DM \lambda} T_{ab}\xi^ak^b\comment{T_{\bm \xi \bm k}} \r]_{\lambda=0} + O(\lambda^2)         \r] \; \DM \Sigma \; \DM \lambda
\end{eqnarray}

This gives
\begin{eqnarray}
O(\lambda^0)&:& \theta(0) = \alpha \l[ T_{ab}\xi^ak^b\comment{T_{\bm \xi \bm k}} \r]_{\lambda=0} 
\label{eq:order-zero}
\\
\nn \\
O(\lambda^1)&:& - \frac{1}{2} \theta^2(0) - \l[ R_{ab}k^ak^b\comment{R_{\bm k \bm k}} \r]_{\lambda=0} = 
\alpha \l[ \frac{\DM}{\DM \lambda} T_{ab}\xi^ak^b\comment{T_{\bm \xi \bm k}} 
\r]_{\lambda=0}
\label{eq:order-lambda}
\end{eqnarray}
Using Eq.~(\ref{eq:order-zero}) to replace $\theta^2(0)$, this becomes
\begin{eqnarray}
O(\lambda^1)&:& - \frac{1}{2} \l[ \alpha T_{ab}\xi^ak^b\comment{T_{\bm \xi \bm k}} \r]^2_{\lambda=0} - 
\l[ R_{ab}k^ak^b\comment{R_{\bm k \bm k}} \r]_{\lambda=0} = \alpha \l[ \frac{\DM}{\DM \lambda} 
T_{ab}\xi^ak^b\comment{T_{\bm \xi \bm k}} \r]_{\lambda=0}
\nn
\end{eqnarray}
The relevant points to note here are:

\begin{itemize}

\item[--] On the horizon, $\kappa \lambda k^a$ goes to $\xi^a$, that is $\l[\kappa \lambda k^a \r]_{\lambda=0} = \xi^a$ (see Appendix \ref{app:wald})
, which is obviously a $O(\lambda^0)$ expression and NOT $O(\lambda)$. This is a key difference from Jacobson's 
argument, arising because Jacobson considers fluxes along generators ${\bar k}^a$ of the horizon. [In that case, 
$\kappa {\bar \lambda} {\bar k}^a=\xi^a$ is valid all across the Killing horizon ($\bar \lambda$ being the affine 
parameter along the generators $\bar k^a$). This then necessitates that expansion of the generators vanish at the 
bifurcation surface $\bar \lambda=0$ (corresponding to $T=0=X$), since the matter flux term becomes $O(\bar \lambda)$.] {In our opinion, since one would expect to associate entropy with cross sections of arbitrary null vectors in an arbitrary curved spacetime, such an assumption on the expansion is restrictive.} 

\item[--] In our {setup}, it would actually be incorrect to deduce that 
$\l[ T_{ab}\xi^ak^b\comment{T_{\bm \xi \bm k}} \r]_{\lambda=0}$ is $O(\lambda)$; as seen from above, this term is 
in fact related to $\theta(0)$, which in general does not vanish. In fact, one would expect arbitrary null congruences to block information of a certain region of spacetime from a class of observers; for such congruences, there is actually no need to constrain the expansion to vanish.

\end{itemize}
So, whether Einstein equations come out at $O(\lambda)$ depends on $\l[ {\DM_\lambda} 
\l( T_{ab}\xi^ak^b \r)\comment{T_{\bm \xi \bm k}} \r]_{\lambda=0} = 
\l[ k^a \nabla_a \l( T_{ab}\xi^ak^b \r)\comment{T_{\bm \xi \bm k}} 
\r]_{\lambda=0}$. In general, it is not at all obvious what this will lead to, but let us 
consider this term in more detail:
\begin{eqnarray}
k^c \nabla_c \l[ T_{ab} \xi^a k^b \r] &=&  \xi^a k^b k^c \underbrace{\nabla_c T_{ab}}_{\mathrm{ignore}} 
\; + \; 
T_{ab} \xi^a \underbrace{k^c \nabla_c k^b}_{= 0}  
\; + \; 
T_{ab} k^b k^c \nabla_c \xi^a
\end{eqnarray}
where we have ignored the derivatives of $T_{ab}$, which is justified in the approximation in which we are working here. For consistency, one must then also ignore the $T_{ab}^2$ term on the LHS of Eq.~(\ref{eq:order-lambda}). Now concentrate on the term involving 
$k^c \nabla_c \xi^a$, which is to be evaluated at $\lambda=0$ {\it after} computing the derivative. This term at 
$\lambda=0$ can be shown to give $-\kappa k^a$\comment{[THIS IS EASILY VERIFIED FOR STATIC SPACETIMES AND IS TRUE IN 
GENERAL]}. Therefore we obtain:

\begin{eqnarray}
k^c \nabla_c \l[ T_{ab} \xi^a k^b \r] \approx  - \kappa T_{ab}k^ak^b\comment{T_{\bm k \bm k}} 
\end{eqnarray}
The last term in $\l[ {\DM} \l(T_{ab}\xi^ak^b\r)\comment{T_{\bm \xi \bm k}} / {\DM \lambda} \r]_{\lambda=0}$ therefore 
reproduces precisely the contribution which would come from calling (in our case incorrectly) 
$\l[ T_{ab}\xi^ak^b\comment{T_{\bm \xi \bm k}} \r]_{\lambda=0}$ as a $O(\lambda)$ term! Therefore the $O(\lambda^1)$ 
becomes [Note that $\alpha = (8 \pi c \lp^2/ \hbar) / \kappa$]:

\begin{eqnarray}
O(\lambda^1) &:& - \l[ R_{ab}k^ak^b\comment{R_{\bm k \bm k}} \r]_{\lambda=0} = - (8 \pi c \lp^2/ \hbar) 
\l[ T_{ab}k^ak^b\comment{T_{\bm k \bm k}} \r]_{\lambda=0}
\nn
\end{eqnarray}
thereby giving the null-null component of Einstein equations yet again, as Jacobson had obtained, but in a 
completely different manner! Also, note that the above relation is applicable all along the future horizon, and not 
just near the bifurcation point, in so far as the notion of local, static Rindler horizon remains well defined. 

The discussion above clearly implies that: 

\begin{itemize}

\item[1.] {In our setup,} Einstein equations do NOT necessarily follow from $T\DM S=\DM E_{\mathrm{matter}}$; this is so since one 
could also have included an additional term, involving curvature tensor, which will only modify the $O(\lambda^0)$ 
term, which is anyway ignored while evaluating the $O(\lambda^1)$ contribution. In fact, as we demonstrated in the 
previous section, such a term is present in Einstein equations, and corresponds to a change in ``gravitational energy". 

\item[2.] The above claims can be explicitly demonstrated by evaluating Einstein equations near a static horizon, 
in which case a very natural energy term is picked out; the resultant equation, in fact, can be thought of as 
$T\DM S - \DM E_{\mathrm{horizon}} = P \DM V = - \DM E_{\mathrm{matter}}$. Note that the $O(\lambda^0)$ 
piece above can be rewritten as $(\kappa/2 \pi) \DM (\delta A/4) = - T_{ab}\xi^ak^b\comment{T_{\bm \xi \bm k}} 
\delta A$, which is, of course, 
Padmanabhan's result without the $\Rtt$ term. Of course, since the latter result is algebraically correct (as we showed in the previous section), this actually makes the $O(\lambda^0)$ contribution (and by implication, the starting relation) incorrect at a \textit{purely algebraic} level -- one needs the $\Rtt$ contribution for consistency [the minus sign above is due to $\bm k$ being an ingoing congruence; as an interesting aside, let us also mention that this fact is one of the ``boundary conditions" for dynamical/isolated horizons imposed by Ashtekar et. al.].

In fact, we feel that the ``$\Rtt$" term reinforces the well known quasi-local character of gravitational energy; an 
ultra-local description of energy balance might therefore be a bit tricky.

\item[3.] Let us also highlight the role of the ingoing congruence $\bm k$. The other null congruence $N^2 \bm l$ 
generates the horizon, and since we have taken great pains to construct a local Killing horizon, the expansion along 
the generators would vanish everywhere on the horizon so long as our local constructs make 
sense; when they don't, we cannot even talk about local Rindler observers and local Killing horizons. Instead, 
the congruence $\bm k$ captures information infinitesimally away from the horizon in the direction normal to it 
[that it is a natural ``normal" is easily seen in inertial coordinates, where future horizon is $T-X=0$ whose 
normal is clearly along $\bm k$.], and provides a natural flow to define variations of various quantities. We hope to 
have shown that it is this congruence which gives, in a sensible manner, the change in a cross-section of the horizon, 
as well as the matter flux across it. Of course, the result can also be justified by applying to \textit{event horizons} 
of a black hole solution of Einstein field equations. For a stationary black hole horizon, the only sensible change in 
area when matter crosses the horizon is given by expansion of $\bm k$, since the expansion of horizon generators vanish 
for a Killing horizon. Moreover, as we have already mentioned before, it has been shown rigorously by 
T. Dray and G. 't Hooft that a massless particle falling into a Schwarzschild black hole corresponds to a shift in 
the ingoing Kruskal coordinate, the shift being proportional to the particle energy. This further strenghtens our 
motivation for using the ingoing congruence for horizon deformations.

\item[4.] One of our main conclusions, as far as comparison with Jacobson's analysis is concerned, is the following:
\textit{The difference between Einstein equations being identical to the first law of thermodynamics, as was pointed 
out by Padmanabhan, differs from the Clausius relation of Jacobson due to difference in the manner matter fluxes 
across the horizon are defined.} Our analysis seems to be closer to the one in Refs. \cite{hayward} 
and \cite{ashtekar-dynamical}.

\item[5.] There are also other significant issues which go beyond the algebraic ones we have mostly concentrated on till now. In using the Clausius relation, one is trying to derive field equations from a starting thermodynamic relation. In such a case, one has to change the starting relation depending on what type of congruence one chooses. These additional terms are then interpreted as dissipation terms and are accounted for by adding suitable entropy production terms in the Clausius relation. However, our analysis above shows that, for a suitably defined horizon (with static near-horizon geometry), the field equations take the form of the first law of thermodynamics without involving any additional terms, under prescribed horizon deformations. The only sensible quantity to concentrate on, while comparing these two approaches, is the $T \DM S$ term (which can have no ambiguity once the entropy density is suitably defined, and which can be verified by applying it to known cases of black hole horizons); in our case, this term is derived using the expansion $\theta$ alone, rather than its derivative, as is required while using Clausius relation. {However, as has already been emphasized several times before, it must be remembered that the difference originates in using different definitions for matter fluxes (and {\it not} due to any in-correctness in any of the two approaces); the physical motivations for the choice used here are mentioned in point (3) above.}

\end{itemize}

Before moving on, we would like to point out that there have been some recent attempts to derive field equations for higher derivative gravity theories \cite{pmb-gen} along the lines of Jacobson. In these works (except Padmanabhan's, see below), the starting point is Clausius relation along with Wald's definition of entropy in terms of Noether charge of diffeomorphism invariance. However, while the definition of matter flux is similar to Jacobson's, the Raychaudhuri equation is never invoked, thereby avoiding any need for assumptions such as vanishing expansion etc. present in original Jacobson work. It would be worth investigating further how the comments in this note are to be considered in the context of these recent attempts. (For one thing, the difference in the definition of matter flux remains.) We must, however, point out that the situation is far from clear since these works do not all agree with each other. For e.g., Parikh and Sarkar have pointed out issues with the Brustein and Hadad paper, whereas Padmanabhan has highlighted certain conceptual issues regarding {\it interpretation} of both these papers. Specifically, Padmanabhan has stressed the subtleties in interpreting these results as derivation of field equations from thermodynamics; the result, he argues, is better viewed as interpretation of field equations as a thermodynamic relation. More importantly, Brustein and Hadad derive their result using a {\it different} form for Noether potential (used to define entropy) than the other two papers; this calls for a more detailed and critical look at the analysis therein. Moreover, the fact that they still derive the same result implies non-uniqueness of the analysis (which a quick look at these papers will confirm), whereas in Jacobson's case, once the assumptions are stated, the analysis is unique. Hence, the status of these results in the light of the original Jacobson calculation remains unclear; the derivations are not only very different from Jacobson's, they are not even similar to each other! Perhaps the most important point indicating why these analyses are {\it conceptually} different from Jacobson's is $f(R)$ theory: whereas the above papers derive the field equations from a Clausius relation, Jacobson and collaborators needed to add extra terms to the Clausius relation in their earlier paper \cite{eling-fR}, to proceed with the analysis. We hope further work will clarify these issues.

\section{Implications}

In this brief paragraph, we would like to emphasize the need for the analysis done in this note. Whether Einstein 
equations are just equations of thermodynamics in disguise is a well motivated question, and we do agree that the 
answer to this might be yes. The important point realized by Jacobson in \cite{jacobson-eq-of-state} while addressing
this question, was to introduce local Rindler frames in an arbitrary curved spacetime, and use the thermal aspects of 
corresponding horizons as probes of the background curvature. However, one needs to impose certain restrictions to 
proceed from there, and to put the result in a physically relevant context. The necessity of highlighting such 
restrictions 
goes hand in hand with identifying specific geometric quantities with (variation of) the thermodynamic variables. 

In 
this sense, as we have shown, the expression for change in entropy of a cross section of a static horizon is related 
to very specific components of the Riemann tensor (and is readily verified for known black hole solutions). Once we 
agree on these algebraic identifications, Einstein equations take the nice form of the first law of thermodynamics, 
provided one attributes to a $2$-surface an energy proportional to its intrinsic curvature. Therefore, Einstein 
equations resemble the first law of thermodynamics in this very specific form. Of course, we have also shown that 
using Raychaudhuri equation also yields, at a higher order (and after justifying ignoring certain terms), the 
null-null component of Einstein equations; the crucial point, however, is that the null-null component of Ricci itself 
has no clear meaning in terms of change of entropy. Looked at in this way, the null-null part of Einstein equations 
seem to be a secondary consequence of the first law itself. Of course, one can reinterpret Einstein equations as 
representing some sort of a Clausius relation; if one insists upon doing so, one must redefine the matter flux 
suitably. It is easy to see that such a definition of flux will involve the trace of the matter stress tensor, and 
will not be equivalent to the heat flux defined by Jacobson; see, for example, reference \cite{hayward} which gives 
one such definition for dynamical horizons, but assuming spherically symmetry. In fact, this can be easily 
demonstrated using our Eq.~(\ref{eq:riemm-area-change}):
\begin{eqnarray}
\l( \frac{\kappa}{2\pi} \r) \frac{\DM ~}{\DM \lambda} \l( \frac{1}{4} \sqrt{\sigma} \r)
&=& \frac{1}{8 \pi} \sqrt{\sigma} \l( R_{ab}k^a\xi^b + \frac{1}{2} R_{\perp} \r) 
\nn \\
&\approx& \sqrt{\sigma} \l( T_{ab} - \frac{1}{2} T g_{ab} \r) k^a\xi^b 
\end{eqnarray}
where $R_{\perp}$ is defined in Appendix \ref{app:riemm-area-change}. We have used the field equations in the second
equality, and the approximation is obtained \textit{after ignoring the $R_{\perp}$ term}. Moreover, inverting the 
logic and deriving field equations in this latter case is again not sraightforward. We hope to have clarified all the above 
issues in the present note.

At a more conceptual level, one of the possible implications of this note is that it might be necessary to adopt a new starting point if one wants to establish Einstein theory in terms of thermodynamics, in which case the thermodynamic structure of the Einstein tensor would serve as the most important supporting evidence. This point of view, of course, also applies to the wider class of Lanczos-Lovelock lagrangians for which similar results hold. A survey of some of the attempts that have been made along these lines can be found in \cite{paddy-newinsights, paddy-equip}.


\section*{Acknowledgements}
I thank T. Padmanabhan for discussions, comments, and careful reading of the manuscript. The author's research is funded by National Science \& Engineering Research Council (NSERC) of Canada, and Atlantic Association for Research in the Mathematical Sciences (AARMS).
\vspace{0.2in}
\appendix

\section{Some mathematical details} \label{app:math-details}

\subsection{Approximate local Killing vector generating Lorentz boosts} \label{app:lif-conds}
The existence of an approximate boost generating Killing field in a locally flat patch of spacetime is a crucial assumption which goes into applying notions of thermodynamics to local acceleration horizons, and it is therefore important to make precise the details of such a construction, which we do below. Around an arbitary event $\mathcal{P}$ in the spacetime, one 
can construct Riemann normal coordinates $y^k$ (with $y^k(\mathcal{P})=0$) in terms of which the metric takes the 
form: $g_{ab} = \eta_{ab} - (1/3)R_{acbd}y^cy^d + O(y^3 \partial R)$. One now wishes to find an approximate Killing 
vector 
field in this patch of spacetime, which would generate approximate Killing boosts along some direction, say $y^1$. 
That is, we want an analog of Minkowski vector field $\xi^M=\kappa[y^1,y^0,0,0]$ where $\kappa$ is acceleration of the 
chosen Rindler observer. This can be done by writing $\xi^k$ as a series in $y^k$ (starting, of course, at linear 
order) and then checking how much freedom one has in the choice of coefficients so as to satisfy Killing equations, 
$S_{ab}=\nabla_a\xi_b+\nabla_b\xi_a=0$. The analysis shows that $\xi^k$ can be made Killing only upto cubic order, 
that is, $S_{ab}=O(y^2)$. Incidentally, this also implies, upon using the identity (valid for arbitrary vector field):
$\nabla_c \nabla_a \xi_b=R_{bacn} \xi^n-(1/2)(\nabla_a S_{bc} + \nabla_c S_{ab} - \nabla_b S_{ca})$, that 
$\nabla_b\nabla_a\xi_c-R_{cabn}\xi^n$ is $O(y)$. We also have: $\Box \xi_b=-R_{bn} \xi^n-\nabla_a \mathcal{S}^a_b$, 
where $\mathcal{S}^a_b=S^a_b - (1/2)S^i_i g^a_b$.

\subsection{Parametrization of null geodesics} \label{app:wald}
[The discussion below mostly follows Wald's book \cite{wald-book}, section 6.4: The Kruskal Extension.]

Consider the simpler form of the Rindler metric: 
$\DM s^2 = - x^2 \DM t^2 + \DM x^2$. The null geodesics satisfy the equation: $\dot t = \pm (\log x)^{\cdot} $, where 
dot denotes derivative w.r.t. affine parameter. Therefore, $t=\pm \log x + $const. for the null geodesics. Further, due to staticity, $-k_0=x^2 \dot t$ is constant along a 
geodesic, so that $\DM \lambda = C x^2 \DM t = \pm C x \DM x$. Hence, $\lambda = \pm C (x^2/2) + D$ is the affine 
parameter. This is the choice we have made. It is further easy to see that the null geodesics are given by 
$k^i=C^{-1} [1/x^2, \pm 1/x]$. Further, note that $C^{-1}$ is the Killing energy associated with the null geodesic; 
imposing $\bm k \cdot \bm \xi = -1$ fixes $C=1$ -- that is, the multiplicative factor in the choice of affine 
parameter is fixed by requiring the null geodesics to have unit Killing energy.

Note that the above parametrization is meaningful only for ingoing geodesics near future horizon (or outgoing 
geodesics near the past horizon). An alternate parametrization is, of course, in terms of the Killing time $t$. It is 
easy to see that this leads to $\lambda \propto \exp{(\pm 2 t)} + $const, which is perhaps a more familiar 
parametrization. Finally, it is easy to deduce the following limits for the contravariant and covariant components of 
the vector $\bm k$: (a) $\left[ x^2 k^i \right]_{x=0} = [1,0] \equiv \xi^i$, 
(b) $\left[ x k_i \right]_{x=0} = [0,\pm 1] \equiv n_i$. These limits clearly confirm that $x=0$ is a null surface, 
and also illustrates the subtlety associated with taking such limits in a coordinate system which is singular (that 
is, when the metric or its inverse blows up in some region, here $x=0$).

\section{Gauss-Codazzi decomposition relations} \label{app:gauss-codazzi-rel}
\noindent \textit{[For greater clarity and notational convenience, we use boldface subscripts for contraction on 
corresponding vectors in this and the following appendices; for e.g., $R_{\bm \xi \bm k} \equiv R_{ab} \xi^{a} k^{b}$.]}

Begin with the Gauss-Codazzi expression for the $4$-$D$ Ricciscalar $R$ with respect to a foliation defined by a 
timelike unit vector $\bm u$:
\begin{eqnarray*}
R = 2 \epsilon_u R_{a b} u^a u^b + \Rt + \epsilon_u \l( K^{\mu \nu} K_{\mu \nu} - K^2 \r)
\end{eqnarray*}
[See, for e.g., \cite{toolkit} sec 3.5.3, page 78.] Similarly, we next decompose the
$3$-$D$ space into a foliation defined by a spacelike vector $\bm n$. This gives
\begin{eqnarray*}
\Rt = 2 \epsilon_n \Rt_{\mu \nu} n^{\mu} n^{\nu} + \Rtt + \epsilon_n \l( k^{A B} k_{A B} - k^2 \r)
\end{eqnarray*}
Above, $\epsilon_u=-1, \epsilon_n=+1$. Similarly, one can use a standard expression to write [with $q_{a b}=g_{ab}+u_a u_b$]
\begin{eqnarray*}
\Rt_{\mu \rho} &=& \Rt_{\mu \nu \rho \sigma} q^{\nu \sigma}
\\
&=& R_{abcd} e^a_{(\mu)} e^c_{(\rho)} \l( g^{bd} + u^b u^d \r) + K^{\nu}_{\mu} K_{\nu \rho} - K K_{\mu \rho} 
\\
&=& R_{ac} e^a_{(\mu)} e^c_{(\rho)} + R_{a u c u} e^a_{(\mu)} e^c_{(\rho)}
+ K^{\nu}_{\mu} K_{\nu \rho} - K K_{\mu \rho}
\end{eqnarray*}
Using this, we obtain
\begin{eqnarray*}
\Rt_{\mu \nu} n^{\mu} n^{\nu} = n^{\mu} n^{\nu} K^{\rho}_{\mu} K_{\rho \nu} - K K_{\mu \nu} n^{\mu} n^{\nu}
+ R_{\bm n \bm n} + R_{\bm n \bm u \bm n \bm u}
\end{eqnarray*}
Putting it all together, we get
\begin{eqnarray}
\Rtt = R &+& 2 \l( R_{\bm u \bm u} - R_{\bm n \bm n} \r) - 2 R_{\bm n \bm u \bm n \bm u} 
\nn \\
&+& f(k)-f(K) - \phi(K)
\label{eq:2Rdecomp}
\end{eqnarray}
where $f(K)=K^2 - K_{\mu \nu}^2$, similarly for $f(k)$, and 
$\phi(K)=n^{\mu} n^{\rho} \l( K^{\nu}_{\mu} K_{\nu \rho} - K K_{\mu \rho} \r)$. Here, $K_{\mu \nu}$ and $k_{AB}$ are extrinsic curvatures of level surfaces of $\bm u$ embedded in 
$4$-$D$ spacetime, and of $\bm n$ embedded in the resultant $3$-$D$ space, respectively. This can be rewritten in a 
better
way by defining $g^{\perp}_{ab}=-u_a u_b + n_a n_b$, and separating out the part depending on extrinsic 
curvatures in 
$\Pi[K,k]=f(k) -  f(K) - \phi(K)$. This yields
\begin{eqnarray}
R = \Rtt + 2 R^{ab} g^{\perp}_{ab} + 2 R_{\bm u \bm n \bm u \bm n} - \Pi[K,k] 
\end{eqnarray}
Note 
that $g^{ab} g^{\perp}_{ab}=2$, $g^{\perp}_{ab} g^{\perp~bc}=g^{\perp~c}_{a}$, and 
$\epsilon_{ab} \epsilon_{cd} = 2 g^{\perp}_{a[d} g^{\perp}_{c]b}$, where $\epsilon_{ab}=2 u_{[a} n_{b]}$ is the 
binormal to the surface.

A slight rearrangement of terms give
\begin{eqnarray}
G_{ab} g^{\perp ab} = - R^{ab} g^{\perp}_{ab} &+& R^{acbd} g^{\perp}_{ab} g^{\perp}_{cd} - \Rtt + \Pi[K,k]
\nn \\
\end{eqnarray}
We stress that the above expression is an identity for any spacelike $2$-$D$ surface embedded in a $4$-$D$ spacetime, and no assumptions such as spherical symmetry or staticity have yet been made. It can be further rewritten as:
\begin{eqnarray}
G_{ab} g^{\perp ab} = \l( R_{\bm u \bm A \bm u \bm B} - R_{\bm n \bm A \bm n \bm B} \r) \sigma^{AB} &-& \Rtt + \Pi[K,k]
\nn \\
\end{eqnarray}
where $R_{\bm u \bm A \bm u \bm B}=R_{abcd} u^a u^c E_{(A)}^{b} E_{(B)}^{d}$, where $E_{(A)}^{a}$ are the dyads 
for the surface. Finally, upon using $\sigma^{AB} E_{(A)}^a E_{(B)}^b = \sigma^{a b} = g^{a b} + u^a u^b - n^a n^b$, 
we arrive at the equation quoted in the text.

\vspace{0.2in}
\section{Derivation of Eq.~(\ref{eq:riemm-area-change})} \label{app:riemm-area-change}
Begin with the identity: 
$R_{\bm \xi \bm k} = - k^a \Box \xi_a = \nabla_c \l( k^a \nabla_a \xi^c \r)$ for $\nabla_{[a} k_{b]} = 0$. Now consider the relation $ k^c \nabla_c \xi^a = - (\partial_z N) k^a $ which is easy to establish for static 
spacetimes. We can then prove the following:
\begin{eqnarray}
\nabla_a \l( k^c \nabla_c \xi^a \r) &=& - (\partial_z N) \nabla \cdot \bm k - k^a \nabla_a (\partial_z N)
\nn \\
&=& - \kappa \nabla \cdot \bm k + O(z^2) + \frac{\partial^2_z N}{N} 
\nn \\
&=& - \kappa  \nabla \cdot \bm k + O(z^2) - \frac{1}{2} R_{\perp}
\end{eqnarray}
where $R_{\perp}$ is the Ricciscalar of the $t-z$ part of the metric: $-N^2 \DM t^2 + \DM z^2$, and is algebraically
equal to $-2 ( \partial^2_z N ) / N$. On the other hand, we also have, from Eq.~(\ref{eq:NEW-exp-area-change}): $\nabla \cdot \bm k = - \frac{\DM ~}{\DM \lambda} \ln \sqrt{\sigma}$.

Putting it all together, we have
\begin{eqnarray}
R_{\bm \xi \bm k} = \kappa \frac{\DM ~}{\DM \lambda} \ln \sqrt{\sigma} - \frac{1}{2} R_{\perp}
\end{eqnarray}

We now note that, for the $2$-$D$ $t-n$ metric, $R_{\bm u \bm n \bm u \bm n} 
= - (1/2) R_{\perp}$ [note that $\bm u$ and $\bm n$ 
form an ortho\textit{normal} basis], which finally yields the desired expression
\begin{eqnarray}
R_{\bm \xi \bm k} - R_{\bm u \bm n \bm u \bm n} = \kappa \frac{\DM ~}{\DM \lambda} \ln \sqrt{\sigma}
\end{eqnarray}
Some aspects of this expression have been discussed in the main text.  We must also mention that the peculiar combination of curvature tensor components above also arises in the analysis in \cite{makela} [see Appendices A and B], in which area variation of an ``acceleration surface" is considered along the observer trajectory [and not along null geodesics as in our case].





\begin{thebibliography}{99}
\bibitem{paddy-newinsights}
T.~Padmanabhan, \textsl{Thermodynamical Aspects of Gravity: New insights}, \textit{Rep. Prog. Phys.} \textbf{73}, 
046901 (2010).
\bibitem{jacobson-eq-of-state}
T.~Jacobson, \textsl{Thermodynamics of Spacetime: The Einstein Equation of State}, \textit{Phys. Rev. Lett.} 
\textbf{75}, 1260 (1995).
\bibitem{paddy-pdv}
T.~Padmanabhan, \textsl{Classical and Quantum Thermodynamics of horizons in spherically symmetric spacetimes}, 
\textit{Class.\ Quantum \ Grav.} \ {\bf 19}, 5387 (2002).
\bibitem{visser}
A.~J.~M.~Medved, D.~Martin, M.~Visser, \textsl{Dirty black holes: Spacetime geometry and near-horizon symmetries}, 
\textit{Class.\ Quantum \ Grav.} \ {\bf 21}, 3111 (2004).
\bibitem{conf-proc}
See sec. (2.1) of: D.~Kothawala, \textsl{Thermodynamic structure of gravitational field equations}, \textit{J.\ Phys.\ : Conf.\ Ser.}\ {\bf 222}, 012014 (2010).
\bibitem{dray-thooft}
T.~Dray, G.~'t Hooft, \textsl{The gravitational shock wave of a massless particle}, 
\textit{Nucl. Phys. B} \ \textbf{253}, 173 (1985). See also, Prof. G. 't Hooft's lecture notes, 
\textsl{Introduction to the theory of Black Holes}, available on: \textit{http://www.phys.uu.nl/~thooft}.
\bibitem{thermod-static}
D.~Kothawala, T.~Padmanabhan, \textsl{Thermodynamic structure of Lanczos-Lovelock field equations from near-horizon 
symmetries}, \textit{Phys. Rev. D} \ {\bf 79}, 104020 (2009).
\bibitem{hayward}
S.~Hayward, \textsl{Unified first law of black-hole dynamics and relativistic thermodynamics}, 
\textit{Class.\ Quantum \ Grav.} \ {\bf 15}, 3147 (1998).
\bibitem{paddy-equip}
T.~Padmanabhan, \textsl{Surface Density of Spacetime Degrees of Freedom from Equipartition Law in theories of 
Gravity}, \textit{Phys. Rev. D} \ {\bf 81}, 124040  (2010).
\bibitem{toolkit}
E.~Poisson, \textsl{A Relativist's Toolkit}, Cambridge University Press, Cambridge (2004).
\bibitem{wald-book}
R.~M.~Wald, \textsl{General Relativity}, University of Chicago Press (1984).
\bibitem{ashtekar-dynamical}
A.~Ashtekar, B.~Krishnan, \textsl{Dynamical Horizons and their Properties}, 
\textit{Phys. Rev. D} \ {\bf 68}, 104030 (2003); 
\textsl{Isolated and Dynamical Horizons and Their Applications}, 
\textit{Living Rev. Relativity} \ {\bf 7}, (2004), 10. \textit{http://www.livingreviews.org/lrr-2004-10}.
\bibitem{pmb-gen}
M.~Parikh, S.~Sarkar, arXiv:0903.1176; T.~Padmanabhan, arXiv:0903.1254; R.~Brustein, M.~Hadad, \textit{Phys. Rev. Lett.} \textbf{103}, 101301 (2009).
\bibitem{eling-fR}
C.~Eling, R.~Guedens, T.~Jacobson, \textsl{Non-equilibrium Thermodynamics of Spacetime}, \textit{Phys. Rev. Lett.} \textbf{96}, 121301 (2006).
\bibitem{makela}
J.~Makela, \textsl{A Simple Quantum-Mechanical Model of Spacetime II: Thermodynamics of Spacetime}, arXiv:0805.3955v3.
\end{thebibliography}
\end{document}